\documentclass[%
 reprint,superscriptaddress,amsmath,amssymb,
aps,nofootinbib]{revtex4-1}
\usepackage{dsfont}
\usepackage{graphicx}

\usepackage{dcolumn}
\usepackage{bm}
\usepackage{color}
\usepackage{xcolor}
\usepackage{epsfig}
\usepackage{soul}
\usepackage[colorlinks=true,urlcolor=brown,linkcolor=blue,citecolor=magenta]{hyperref}
\usepackage{natbib}
\usepackage{float}
\usepackage{footmisc}
\usepackage{siunitx}
\usepackage[normalem]{ulem}

\definecolor{lightred}{rgb}{1,0.5,0.5}
\definecolor{lightgreen}{rgb}{0.5,1,0.5}
\definecolor{lightblue}{rgb}{0.5,0.5,1}
\definecolor{lightcyan}{rgb}{0.5,0.75,0.75}
\definecolor{lightmagenta}{rgb}{0.75,0.5,0.75}
\definecolor{customgreen}{rgb}{0.494,1,0.502}

\newcommand{\meV}{\mathinner{\mathrm{meV}}}

\newcommand{\MeV}{\mathinner{\mathrm{MeV}}}
\newcommand{\GeV}{\mathinner{\mathrm{GeV}}}
\newcommand{\TeV}{\mathinner{\mathrm{TeV}}}

\newcommand{\meters}
{\mathinner{\mathrm{m}}}

\begin{document}

\title{
\framebox[14cm]{\rule[-.85cm]{0cm}{20mm} 
\begin{tabular}{c}
 Searching for a new light gauge boson with axial couplings
 \vspace{3mm} \\
in muon beam dump experiments
\end{tabular}}
\vspace{6mm}
}

\author{Pierre Fayet} 
\email{pierre.fayet@phys.ens.fr}
\affiliation{Laboratoire de physique de l'École normale supérieure, ENS, Université PSL, CNRS, Sorbonne Université, Université Paris Cité, F-75005 Paris, France}
\affiliation{Centre de physique th\'eorique, \' Ecole polytechnique, IPP, 91128 Palaiseau cedex, France}

\author{María Olalla Olea-Romacho\vspace{5mm} }
\email{mariaolalla.olearomacho@phys.ens.fr}
\affiliation{Laboratoire de physique de l'École normale supérieure, ENS, Université PSL, CNRS, Sorbonne Université, Université Paris Cité, F-75005 Paris, France}

\begin{abstract}
\vspace{8mm}

We present a formalism for new $U(1)$ interactions involving weak hypercharge, baryon, and lepton numbers, and a possible axial symmetry generator $F_A$ in the presence of a second Brout-Englert-Higgs doublet. The resulting $U$ boson, after mixing with the $Z$, interpolates between a generalised dark photon, a dark $Z$, and an axially coupled gauge boson. We especially focus on the axial couplings originating from $F_A$ or from mixing with the $Z$, determined by the scalar sector via parameters like $\tan\beta$ and the v.e.v.~of an extra dark singlet.

\vspace{1mm}
We explore the distinctive features of axially coupled interactions, especially in the ultrarelativistic limit, where the $U$ boson behaves much as an axion-like particle, with enhanced interactions to quarks and leptons. This enhancement is particularly relevant for future muon beam dump experiments, since the muon mass considerably increases the effective coupling, proportional to $2m_\mu/m_U$, compared to analogous experiments with electrons.

\vspace{1mm}
We also analyse the shape of the expected beam dump exclusion or discovery regions, influenced by $U$ boson interactions and the experiment geometry. Different situations are considered, limited in particular by cases for which the $U$ decays before reaching the detector, or has too small couplings to produce detectable events. We also compare to vectorially coupled bosons and axion-like pseudoscalars, highlighting the importance of understanding the parameter space for future experiment design and optimisation.
\vspace{12mm}

\end{abstract}

\vspace{8mm}

\maketitle

\section{Introduction}
In addition to the weak, electromagnetic and strong interactions mediated by $W^\pm$ and $Z$ bosons, photons and gluons, there may well be new interactions. The corresponding bosons, which may be associated with extra $U(1)$ factors in the gauge group, could be relatively light or even very light, provided they are sufficiently weakly coupled~\cite{Fayet:1980ad,Fayet:1980rr,Fayet:1990wx}. Such gauge bosons, denoted by $U$, appear as generalised dark photons coupled to a linear combination of the standard model symmetry generators, specifically $Q$ with $B$ and $L_i$ (or $B-L$). Axial couplings can be present when the electroweak symmetry is broken by two spin-0 Brout-Englert-Higgs (BEH) doublets, alongside an additional dark singlet. A light $U$ boson may also induce sufficient annihilations to allow for light dark matter particles, providing a bridge between standard model particles and a new dark sector~\cite{Boehm:2003hm,Fayet:2004bw}. This has led to many theoretical and experimental studies, and there is now a considerable interest in searching for such a dark sector~\cite{Bossi:2012uwa,Battaglieri:2017aum,Fabbrichesi:2020wbt,Graham:2021ggy,Antel:2023hkf}.

\vspace{1.5mm}

We shall discuss here how muon beam dump experiments can contribute to these searches, and also illustrate how a light boson with axial couplings undergoes somewhat {\it enhanced\,} interactions for its longitudinal polarisation state, 
compared to its transverse ones. In fact, the production cross section of a longitudinal $U$ boson with axial couplings $g_A$ is proportional  to $g_A^2/m_U^2\propto\ $[extra-$U(1)$ breaking scale]$^{-2}$, as for an axion-like pseudoscalar. This feature makes its phenomenology different from that of a spin-1 boson with only vector interactions. The difference is especially important for projected muon beam dump experiments because the larger mass of muons significantly boosts the effective interaction strength, proportional to $2m_\mu/m_U$, compared to electron beam dump experiments.

\vspace{1.5mm}

A second effect, in the opposite direction, is that the strength of these effective pseudoscalar interactions can be significantly {\it reduced\,} when the extra $U(1)$ symmetry is broken sufficiently above the electroweak scale by a large dark singlet v.e.v.~This makes the nearly equivalent axion-like pseudoscalar mostly an electroweak singlet, and thus largely ``invisible''. Its production and interaction amplitudes are then reduced by an invisibility factor $r < 1$, and its exchange amplitudes by $r^2$ \cite{Fayet:1980ad,Fayet:1980rr}.

\vspace{1.5mm}

Both types of effects can  also be present in locally supersymmetric theories, where the  $\pm \,1/2$ polarisation states of a very light spin-3/2 gravitino 
undergo {\it enhanced\,} interactions, compared to the gravitational interactions of the $\pm \,\frac{3}{2}$ polarisation states, 
\vspace{-.3mm}
proportional to 
$\,G_N/m_{3/2}^2\propto$ 
\vspace{-.6mm}
[supersymmetry-brea\-king scale]$^{-2}$~\cite{Fayet:1977vd}. Such a light gravitino behaves very much as a spin-1/2 goldstino, with potentially sizeable interactions if the supersymmetry-breaking scale is comparable to the electroweak scale.
These effective interactions, however, get {\it reduced\,} if supersymmetry is broken at a large scale ($\sqrt d$ or $\sqrt F)$, significantly larger than the electroweak scale, through the large v.e.v.~of an auxiliary field. The almost-equi\-valent goldstino then behaves as a quasi-``invisible'' particle, just as a longitudinal $U$ boson behaves as a quasi-invisible axion-like pseudoscalar, if the corresponding $U(1)$ symmetry is broken sufficiently above the electroweak scale by a large dark-singlet v.e.v.

\vspace{1.5mm}

In this paper, we shall concentrate on an extra gauge boson featuring axial interactions, associated with a second BEH doublet  allowing for an additional axial symmetry generator $F_A$. Taking into account mixing effects with the $Z$ boson, we shall discuss how the axial couplings $g_A$, the mass $m_U$ and the invisibility parameter $r$ affect its production and detection in a muon beam dump experiment, and the resulting limits that can then be obtained. We shall investigate: 1) the axion-like behaviour of an axially coupled light gauge boson, with enhanced interaction strength with quarks and leptons, compared to its vectorially coupled equivalent; 2) the impact of the parameters of the scalar sector on the $U$ boson phenomenology. For example, for a given coupling constant $g_A$ and value of the ratio of the two doublet v.e.v.s, $\tan \beta$, the mass of the $U$ boson cannot be arbitrarily small, since two BEH doublets are required to gauge an axial symmetry, imposing a minimum mass limit. This previously unaddressed aspect arises as a natural consequence of considering axially coupled interactions; 3) the shape of the beam dump exclusion or discovery region, determined by the interplay of the $U$ boson interactions and the experiment geometry. This analysis highlights the importance of understanding the parameter space and the interaction dynamics for the optimal design of future beam dump experiments.

\vspace{1.5mm}
In sections \ref{sec:Mass_U} and \ref{sec:couplings_U}, we provide a general discussion on the mass and couplings of a $U$ boson within a two-Higgs-doublet model (2HDM) extended with a complex scalar singlet. This includes an analysis for both the axial and vector couplings of the $U$ boson with a special emphasis on the former. In sections \ref{sec:enhanced_int} and \ref{subsec:axial} we discuss the special behaviour of the longitudinal polarisation state of a light $U$ boson with axial couplings to standard model fermions, mimicking the interactions of a quasi-invisible axion-like particle. In section~\ref{sec:beam_dump}, we study the production, signature and phenomenology of an axially coupled $U$ boson in muon beam dump experiments. Finally, we conclude in section~\ref{sec:conclu}.

\section{Mass and \vspace{2mm}couplings \hbox{of a new light gauge boson}}

\subsection{\boldmath Mass mixing with the $Z$}
\label{sec:Mass_U}

The new gauge boson may have in general axial couplings, if two BEH doublets participate in the electroweak symmetry breaking. These couplings may originate from the presence of an axial $U(1)_A$ factor in the gauge group, possibly with a very small gauge coupling $g''$,
\footnote{\label{susy} This axial $U(1)_A$ was originally considered within supersymmetric theories, 
in which it acts on left-handed chiral superfields according 
\vspace{-1mm}to
$$
(H_1,\!H_2) \to e^{i\alpha}\,(H_1,\!H_2),\  (Q,\bar U,\bar D;L,\bar E) \to e^{-i\alpha/2}
(Q,\bar U,\bar D;L,\bar E),
$$
so that the trilinear superpotential responsible for quark and lepton masses is left invariant. Its definition is extended to an extra singlet $S$ interacting with the two doublets $H_1$ and $H_2$ through the trilinear superpotential $\lambda \,H_2H_1 S$, according to
$S \to e^{-2i\alpha}\,S$ \cite{Fayet:1974pd,Fayet:1977yc}. Its spontaneous breaking, assuming anomalies to be cancelled, or irrelevant due to a very small gauge coupling $g''$, generates a massless axion-like Goldstone boson, ``eaten away'' when the gauge boson acquires a mass.
} and/or from a mixing between the extra-$U(1)$ gauge field $C^\mu$ and the $Z^\mu$ field of the standard model. 
\vspace{1.5mm}

We shall generally assume that $h_1$ (with weak hypercharge $Y=-\,1$) is responsible for down-quark and charged-lepton masses, and $h_2$ (with $Y=+\,1$) for up-quark masses, as in supersymmetric theories (or as in type II 2HDMs). The analysis also applies to other situations, including type I 2HDMs, in which a single doublet $h_1$ is responsible for all quark and lepton masses. The covariant derivative $iD_\mu$ is expressed as
\begin{equation}
\label{dcov}
iD_\mu\,=\,i\partial_\mu - g \ \hbox{\boldmath $T.W$}_{\!\mu}-\frac{g'}{2}\,Y \,B_\mu- \frac{g''}{2}\,F\,C_\mu\,,
\end{equation}
ignoring the QCD term, which is not relevant here. $F$ denotes the quantum number associated with the extra $U(1)$ gauge group. Without loss of generality, we also disregard a possible kinetic-mixing term between $U(1)$ gauge fields. Such a term is not present in an orthogonal field basis, and is otherwise easily removed by diagonalisation. 

\vspace{1.5mm}

Mixing effects with the $Z$ boson arise after electroweak symmetry breaking, and are independent of how quark and lepton masses are generated. Let $F_{1,2}$ represent the additional $U(1)$ quantum numbers associated with each one of the BEH doublets $h_{1,2}$. The $(2\times 2)$ mass-squared matrix ${\cal M }^2$, derived from eq.\,(\ref{dcov}) in the $(Z_{\rm sm}^\mu, C^\mu)$ basis, is expressed \vspace{2mm}as follows
\begin{equation}
\label{m2}
\frac{1}{4}\,\left[\begin{array}{cc}
  g_Z^2\,(v_1^2+v_2^2)       & \  g''g_Z (F_1v_1^2-F_2v_2^2) \vspace{3mm}\\
   g''g_Z (F_1v_1^2-F_2v_2^2)     & \ g''^2(F_1^2v_1^2+F_2^2v_2^2+ F_\sigma^2 w^2)
    \end{array} \right],
\end{equation}
\vspace{-1mm}

\noindent
where $\langle h_i^0 \rangle = v_{i}/\sqrt 2$\,.
Here $g_Z = \sqrt{g^2+g'^2}\,$, 
\vspace{.3mm}
and $\,Z_{\rm sm}^\mu= $ $\cos\theta \ W^\mu_3 - \sin\theta \,B^\mu$ is the usual expression of the $Z$ field in the standard model. Furthermore, $\tan\beta = v_2/v_1$ 
\linebreak
is the ratio of the two \hbox{spin-0} doublet v.e.v.s, 
\vspace{-.6mm}
so that
$v_1=v\cos\beta$, and $v_2 = v\sin\beta$ 
with $v = 2^{-1/4}\,G_F^{-1/2}\simeq \,246$ GeV.
We have also included
the contribution $\,g''^2\,F_\sigma^2\,w^2/4$ associated 
\vspace{-.3mm}
with an extra dark singlet of 
$U(1)$ charge $F_\sigma$ and v.e.v.~$\langle \sigma \rangle = \,w/\sqrt 2$, which could be alternatively replaced by a direct mass term for $C^\mu$.
\vspace{2mm}

This leads to a $Z$-$U$ mixing angle $\xi$ corresponding to the new physical fields
\begin{equation}
\left\{
\begin{array}{ccc}
 Z^\mu \!&=& \cos\xi \ Z^\mu_{\rm sm}  -\sin\xi \ C^\mu\,,  
 \vspace{2mm}\\
 U^\mu \!&=& \sin\xi \ Z^\mu_{\rm sm}  +\cos\xi \ C^\mu\,,  
\end{array}
\right.
\label{eq:fields}
\end{equation}
while the expression of the photon field $\,A^\mu = \sin\theta \ W^\mu_3 + \cos\theta \,B^\mu\,$
is left unchanged with respect to the standard model. The mixing angle $\xi$, which is small in the context of the small $g''$ values considered in this paper, is derived from eq.\,(\ref{m2}) and given by~\cite{Fayet:1986rh,Fayet:1990wx}
\begin{equation}
\begin{array}{ccl}
\label{xi}
\tan\xi &\simeq &\displaystyle \frac{g''}{g_Z}\ (F_2\, \sin^2\beta-F_1\, \cos^2\beta )\,,

\end{array}
\end{equation}
whereas the $U$ boson mass is given by
\begin{equation}
\label{mu-1}
  \!\!m_U\simeq \,\frac{g''\cos\xi}{2} \, \sqrt{\,\hbox{\small$\displaystyle \left(\frac{F_1+F_2}{2}\right)^2$}\sin^2 2\beta\ v^2 + F_\sigma^2 \,w^2}\,.
\end{equation}
\vspace{1mm}

When $F_1+F_2$ does not vanish, so that $h_1$ and $h_2^c$ have different gauge quantum numbers, we can normalise $g''$ so that $(F_1+F_2)/2 = 1$. Then, we have 
\begin{equation}
\label{mu0}
\framebox[7cm]{\rule[-.35cm]{0cm}{10mm} \hbox{$ \displaystyle
  m_U\,\simeq \,\frac{g''\cos\xi}{2} \ \sqrt{\,\,\sin^2 2\beta\, v^2 + F_\sigma^2 \,w^2}$}\ ,}
\end{equation}
with e.g.~$F_\sigma = -\,2$, as found in theories inspired by supersymmetry (cf.~footnote~\ref{susy}).
\vspace{1.5mm}

On the other hand, when $F_1= -F_2$, so that $h_1$ and $h_2^c$ (both with $Y=-1$) share the same gauge quantum numbers, the gauge boson squared-mass matrix $ {\cal M }^2$ in eq.\,(\ref{m2}) coincides with the one arising from a single doublet $h$ with v.e.v.~$v/\sqrt 2$, 
leaving an unbroken $U(1)_U$ gauge symmetry associated with a massless $U$ boson. This residual $U(1)$ symmetry gets broken by the v.e.v.~of the additional dark singlet $\sigma$, providing a small mass for the $U$ boson, 
\begin{equation}
m_U\,\simeq \ \frac{g''\cos\xi}{2}\ |F_\sigma| \,w\,.
\end{equation}
Here we recover the usual situation of a dark photon, or generalised dark photon coupled to a combination of $Q$, $B$ and $L_i$. 

\subsection{\boldmath The couplings of the $U$ boson}
\label{sec:couplings_U}
In this section, we recall how the $U$ boson arising from a new $U(1)$ interaction interpolates between a generalised dark photon coupled to $Q$, $B$ and $L_i$, a dark $Z$ boson coupled to the $Z$ current, and a gauge boson axially coupled to quarks and leptons.
Since we intend to make a special emphasis on the axial couplings of the new interaction, we will specify in which cases it is possible for an extra axial $U(1)_A$ symmetry to participate in the gauging. This requires two BEH doublets involved in the generation of quark and lepton masses, as in supersymmetric theories or type II 2HDMs. The axial $U(1)_A$ symmetry generator $F_A$ is equal to $+\,1$ for $h_{1,2}$, and $\mp\,\frac{1}{2}$ for left-handed and right-handed quark and lepton fields, respectively (cf.~footnote~(\ref{susy})). We can express the extra $U(1)$ generator $F$ for the standard model fields by combining the weak hypercharge $Y$ with the axial generator $F_A$, and baryon and lepton numbers $B$ and $L_i$, according to
\begin{equation}
\label{F}
F\,=\,\gamma_A \,F_A + \gamma_Y \,Y +\gamma_B \,B + \gamma_i \,L_i\ .
\end{equation}
$\gamma_A$ and $\gamma_Y$ are then related to the $F$ quantum numbers  of
$h_1$ and $h_2$ by
\begin{equation}
\label{FF}
F_1=-\,\gamma_Y + \gamma_A\,,\ F_2=\,\gamma_Y + \gamma_A\,.
\end{equation} 

\vspace{1mm}

The new $U$ boson, described by the field in eq.\,(\ref{eq:fields}),
appears as a generalised dark photon coupled 
to 
\begin{equation}
(g''/2)\,F\, \cos\xi + g_Z \,Q_Z\,\sin\xi \, , 
\end{equation}
where $g_Z =\sqrt{g^2+g'^2}$\,, $Q_Z =T_{3L}-\sin^2\theta \, Q$, $Q= T_{3L}+Y/2$ being the electric charge and $\theta$ the electroweak mixing angle.
It is convenient to re-express the small $Z$-$U$ mixing angle $\xi$ in eq.~(\ref{xi})
so that
\begin{equation}
\label{xi2}
\tan\xi \,\simeq \frac{g''}{g_Z}\ (F_2\, \sin^2\!\beta-F_1\, \cos^2\!\beta )\,
\,= \,\frac{g''}{g_Z}\, (\gamma_Y+\eta)\,.
\end{equation}
This implies from eq.~(\ref{FF})
\begin{equation}
\label{eta}
\eta\,=\,-\,\gamma_A\,\cos 2\beta\,.
\end{equation}
Independently of the specific expression of $\eta$, the $U$ boson is coupled 
with strength $g''\cos\xi $ to the following $U$ charge, expressed for standard model fields in the limit of small $g''$ as~\cite{Fayet:2020bmb}
\begin{equation}
\label{qu}
\begin{array}{ccl}
 Q_U    &= & \frac{1}{2}\,F\,+\,\tan\xi\ \, {\displaystyle \frac{g_Z}{g''}}\ (T_{3L} - \sin^2\theta \,Q)  \vspace{2mm}\\
 &\simeq & \,\frac{1}{2}\,(\gamma_A \,F_A +\gamma_Y\,Y + \gamma_B \,B + \gamma_i \,L_i) \vspace{3mm}\\
     & &\ \ \ \ +\ (\gamma_Y +\eta) \ (T_{3L} - \sin^2\theta \,Q)\ .
\end{array}
\end{equation}

Altogether,
\begin{equation}
\label{qu2}
\framebox[8cm]{\rule[-.6cm]{0cm}{15mm} 
\hbox{$
\begin{array}{ccl}
\ Q_U  \! \! &= &\! \gamma_Y\cos^2\theta \ Q \,+ \,\frac{1}{2}\,(\gamma_A \,F_A +\gamma_B \,B + \gamma_i \,L_i)\vspace{3mm}\\
     & &\ \ \ \ +\ \eta \ (T_{3L} - \sin^2\theta \,Q)\,,\
\end{array}
$}
}
\end{equation}
 shows the various aspects of the new $U$ boson, indicating that it may appear as a pure dark photon, a dark $Z$, an axial gauge boson, a gauge boson for $B$ and $L_i$ (or $B-L$ or $L_i-L_j$), or that it generally interpolates between all these possible aspects.

\vspace{2mm}

Let us now distinguish the vector from the axial couplings of the $U$ boson. Considering all coefficients $\gamma_A,\,\gamma_Y,\,\gamma_B,\,\gamma_{L_i}$ and $\eta$ in the $Q_U$ charge (see eq.\,(\ref{qu2})),
the vector couplings of the $U$ boson are obtained as a general linear combination of $Q$ with $B$ and $L_i$ (or $B-L$ in a grand unified theory), reexpressed in a general way as 

\vspace{-5mm}

\begin{equation}
(\varepsilon_Q\,Q + \varepsilon_B B +\varepsilon_i\,L_i) \times e\ .
\end{equation}

\vspace{1mm}

Next we turn to the axial couplings, which are also obtained from the same general formula, eq.~(\ref{qu2}). With left-handed and right-handed projectors expressed as 
\hbox{$(1\mp \gamma_5)/2$} and $F_A = \mp\, 1/2$ for left-handed and right-handed quark and lepton fields, respectively, the axial couplings read 
\begin{equation}
\label{gapm-1}
g_{A\pm}\, \simeq \ \frac{g''}{4}\,\cos\xi\ ( \gamma_A \mp \eta) \,.
\end{equation}
$g_{A+}$ refers to up-type quarks and neutrinos, and
$g_{A-}$ to down-type quarks and charged leptons.
The isoscalar contribution to the axial couplings originates from the participation of the axial generator $F_A$ in the gauging, and the isovector part from the supplementary contribution from the $Z$-$U$ mixing, as measured by the parameter $\eta$ in eqs.\,(\ref{xi2},\ref{qu2},\ref{gapm-1}). 

\vspace{1.5mm}
For $\gamma_A\neq 0$, and by normalising $g''$ so that $\gamma_A =1$, the axial couplings
are expressed from eqs.~(\ref{eta},\ref{gapm-1}) as
\begin{equation}
\label{gapm}
\framebox[5.4cm]{\rule[-.3cm]{0cm}{8mm}\hbox{$
\displaystyle \ g_{A\pm} \,\simeq \,\frac{g''}{4}\,\cos\xi\ ( 1 \pm \cos 2\beta)\,.
$}
}
\end{equation}
In particular, we have for down-type quarks and charged leptons, including the muon,
\begin{equation}
\label{gamu}
\displaystyle g_{A\,d,e}\,\simeq\,\frac{g''}{2}\,\cos\xi\, \sin^2\!\beta\,, 
\end{equation}

\noindent
where $\cos\xi\simeq 1$ for a very light $U$ boson. The axial coupling $g_{A_{d,e}}$ reduces to $g''/4$ when $v_1 = v_2$, a condition under which no additional mixing with the $Z$ boson occurs as shown in eq.~(\ref{xi2}) (that is, $\eta = 0$, even if $\gamma_Y \neq 0$). Consequently, $Q_Z$ is then absent from the $Q_U$ expression in eq.~(\ref{qu2}).
\vspace{2mm}

\paragraph*{Dark photon case.} One  recovers the pure dark photon situation by choosing $F_1 = - F_2$, so $h_1$ and $h_2^c$ jointly behave as a single doublet. Alternatively, one can also consider a single doublet $h$ as in the standard model, so $F_A$ does not take part in the gauging. 
In this case, by setting $\gamma_A=\gamma_B=\gamma_{L_i}= \eta =0$, and normalising $g''$ to $\gamma_Y=1$, we get $Q_U = \cos^2\theta\, Q$ as a result of the $Z$  mixing with the $U$. The $U$ boson appears as a pure dark photon, with a coupling of
\begin{equation}
\varepsilon_Q \,e\, \simeq \,g''\cos\xi\,\cos^2\theta
\label{eq:DP_case}
\end{equation}
to the electric charge. This is obtained from the participation of $Y$ in the gauging of the extra $U(1)$ symmetry, without having had to consider any kinetic-mixing term~\cite{Fayet:1990wx}. 
\vspace{1.5mm}

\vspace{2mm}

\paragraph*{Type I 2HDM's.} As mentioned above, we cannot define 
 an axial quantum number $F_A$ if only one of the BEH
doublets is responsible for both the up-type and down-type quark masses.
By fixing $\gamma_A = 0$, the expression in eq.~(\ref{qu2}) still applies if all quarks and leptons receive their masses from the same doublet (for instance, $h_1$).
This does not affect down-type quarks and charged leptons, that still receive their masses from $\langle h_1\rangle$, resulting in the same $g_{A\,d,e}$ as in eq.~(\ref{gamu}). We derive the isovector axial couplings by setting $F_1 = -\,\gamma_Y$ and $F_2 = \gamma_Y + 2$, and keeping $g''$ normalised so that $(F_1 + F_2)/2 = 1$. The expressions~(\ref{mu-1},\ref{mu0}) for $m_U$ remain valid as well. Together with $\gamma_A =0$ and $\eta= 2 \sin^2\!\beta$, we get from eqs.~(\ref{xi2},\ref{qu2},\ref{gapm-1}) 
\begin{equation}
\label{gapm2}
g_{A\pm} \,\simeq \ \mp \ \frac{g''}{4}\, \cos\xi\  \eta \ \simeq \ \mp  \ \frac{g''}{2}\, \cos\xi\, \sin^2\!\beta\,,
\end{equation}
which leads to the same axial couplings for down-type quarks and charged leptons, specifically for the muon, as for type II 2HDMs (see eqs.~(\ref{gamu})).

\subsection{\boldmath The axion-like behaviour of the $U$ boson
}
\label{sec:enhanced_int}

An essential feature in the presence of axial couplings is that the interaction amplitudes for a new light gauge boson are enhanced by a factor $\propto k^\mu/m_U$ for its longitudinal polarisation state.
It then effectively behaves as a nearly equivalent pseudoscalar particle, with effective pseudoscalar couplings to quarks and leptons~\cite{Fayet:1980rr}
\begin{equation}
\label{gpga}
\framebox[3.2cm]{\rule[-.35cm]{0cm}{9mm} \hbox{$ \displaystyle
  g_p\, =\, g_A\ \frac{2m_{q,l}} {m_U} \,.
  $}
  }
\end{equation}

\vspace{1mm}
 In the presence of a dark singlet $\sigma$ with
v.e.v.~$\langle \sigma\rangle = w/\sqrt 2\,$, $m_U$ increases from a value $m_U^0 = g'' v \,\sin 2\beta \,/2$ (induced only by the two BEH doublets v.e.v.s.) up to the value specified in eq.~(\ref{mu0}). This increase in the mass of the $U$ boson, quantified by the factor
\begin{equation}
\label{rmu}
    \frac{1}{r}\,=\,\frac{m_U}{m_U^0}\,=\,\frac{\sqrt{\,v^2\sin^2 2\beta+ F^2_\sigma\, w^2}}{v\,\sin 2\beta}\,>\, 1\,,
\end{equation}
 is accompanied by a corresponding decrease in the production and interaction amplitudes of its longitudinal polarisation state. These are proportional to $1/m_U$ or, equivalently, to $r$~\cite{Fayet:1980ad,Fayet:1986rh} 
\begin{equation}
\label{rinv}
\framebox[5.9cm]{\rule[-.5cm]{0cm}{11mm} \hbox{$ \displaystyle
    r\,= \, \frac{v\,\sin 2\beta}{\sqrt{\,v^2\sin^2 2\beta+ F^2_\sigma\, w^2}}\,<\, 1\,.$}}
\end{equation}

\noindent
 More precisely, the relevant quantity to measure the effective strength of the interactions of a longitudinal $U$ boson with both up-type and down-type fermions, as we shall see soon from eqs.\,(\ref{gpax},\ref{gpax2}), is
\begin{equation}
    r\,\tan\beta + r\,\cot\beta \,= \, \hbox{$\displaystyle \frac{2v}{\sqrt{\,v^2\sin^2 2\beta+ F^2_\sigma\, w^2}} $}\ .
\end{equation}
The invisibility parameter is much smaller than 1 when $|F_\sigma|\, w$ is much larger than the electroweak scale, and the extra $U(1)$ symmetry is broken at a higher energy scale. In this case,  the interactions of a longitudinal $U$ boson can become arbitrarily small.
\vspace{2mm}

In fact, such a longitudinal $U$ boson behaves much like an effective pseudoscalar $a$, the Goldstone boson that is ``eaten away" when $U$ acquires its mass.
$a$ is then mostly an electroweak singlet close to $\sqrt 2$ Im $\sigma$, 
\vspace{.1mm}
in the presence of a large singlet v.e.v., and thus nearly ``invisible''. 
More precisely, this pseudoscalar field $a$ is a combination of $\sqrt 2$ Im $\sigma$ and the $CP$-odd field  $A=\sqrt 2\ \hbox{Im}\,(\sin\beta\ h_1^0 + \cos\beta\ h_2^0)$,
 which is orthogonal to the combination 
 \vspace{-.4mm}
$z_g = \sqrt 2\ \hbox{Im}\,(\cos\beta\ h_1^0 - \sin\beta\ h_2^0)$ 
eaten away by the $Z$ boson. It thus reads, independently of $\gamma_Y$ and of the specific Yukawa couplings of $h_1$ and $h_2$ to quarks and leptons,
\begin{equation}
\label{inva}
\!\!
 a = \sqrt 2\ \,\hbox{Im}\ [\,\cos\theta_A \,(\sin\beta\,h_1^0 + \cos\beta\,h_2^0) +\, \sin\theta_A \,\sigma\,].\!
\end{equation}
The production and interaction amplitudes for a longitudinal light $\,U$ boson, which behaves as a mostly-singlet axion-like pseudoscalar, are then reduced by the invisibility factor

\vspace{-4mm}
\begin{equation}
r \,=\,\cos\theta_A\,,
\end{equation}
already expressed in eq.~(\ref{rinv}).

\vspace{1.5mm}

The pseudoscalar Yukawa couplings 
\vspace{-.3mm}
of $\sqrt 2$ Im $h_1^0$ and $\sqrt 2$ Im $h_2^0$ are
given by $m_f/ (v\cos\beta)$ and $m_f/ (v\sin\beta)$, respectively, if $h_1$ and $h_2$ are separately responsible for down-quark and charged-lepton masses, and up-quark masses. In this case,
expression (\ref{inva})  leads to effective pseudoscalar couplings
\begin{equation}
\label{gpax}
\frac{m_f}{v} \times \left\{
 \begin{array}{cl}
r\,\cot\beta &\hbox{for up quarks},     \vspace{1mm} \\
r\,\tan\beta & \hbox{for down quarks and charged leptons}.
\end{array}\right.    \end{equation} 

\vspace{1mm}
\noindent

Eq.~(\ref{inva}) for $a$ still remains valid even if only $h_1$, for instance, is responsible for all standard model fermion masses, as in type I 2HDMs. This situation, for which $F_A$ does not participate in the gauging, results in effective pseudoscalar couplings $\,\mp \,(m_f/v) \,\times \,r \,\tan \beta\,$ for all standard model fermions.

\vspace{2mm}

The above picture, based on a model with two doublets and one complex singlet, relies on a $U(1)_A$ or an additional $U(1)$ symmetry broken significantly above the electroweak scale through a large singlet v.e.v., which results in a substantial reduction of the interaction amplitudes by the factor $r=\cos\theta_A$. The similarity between the interactions of a light gauge boson with axial couplings and those of the corresponding axion-like pseudoscalar, considered in the presence of a large singlet v.e.v., also led us to independently propose in 1980~\cite{Fayet:1980ad} the invisible axion mechanism~\cite{Dine:1981rt,Zhitnitsky:1980tq}.
\vspace{2mm}

The axion field, expressed very much as in eq.~(\ref{inva}), is then predominantly an electroweak singlet, with a small contamination, proportional to $\cos\theta_A$, arising from the doublet components associated with electroweak symmetry breaking. This is well illustrated by the expressions of the branching ratios
\begin{equation}
\label{gpax2}
\!\!\!\left\{
\begin{array}{ccc}
B(\psi \to\gamma\ U/a)\! &\!\propto\!&\! (r=\cos\theta_A)^2 \times (x = \cot \beta)^2\,,
\vspace{2mm}\\
B(\Upsilon \to\gamma\ U/a)\!&\! \propto\!&\! (r=\cos\theta_A)^2 \times \displaystyle(\frac{1}{x} = \tan \beta)^2\,,
\end{array} \right.\!\!\!
\end{equation}
obtained by rescaling the branching ratios for a standard axion \cite{Wilczek:1977pj,Weinberg:1977ma} by the invisibility factor $r$ in eq.~(\ref{rinv}). 
Experiments with $\psi$ and $\Upsilon$ decays have long since ruled out the possibility of $r=1$, indicating the need for an additional singlet with sufficiently large v.e.v. See e.g. ref.~\cite{Fayet:2020bmb} for a discussion of various experimental constraints on the invisibility parameter $r$.

\subsection{\boldmath \vspace{1mm} Equivalent pseudoscalar couplings}

\label{subsec:axial}

The mechanism providing reduced interactions for the $U$ boson is associated with the increase of the $U$ boson mass from the dark singlet contribution in eq.~(\ref{mu0}), leading to
\vspace{-5mm}

\begin{equation}
\label{mu}
 m_U    \, \simeq \, \displaystyle\frac{g''v\,\cos\xi}{2}\ \frac{\sin 2\beta}{r}\,.
\end{equation}
It allows us to express the extra-$U(1)$ gauge coupling $g''$, and thus the axial couplings $g_A$ in eq.~(\ref{gapm-1}), as proportional to both $m_U$ and $r$, according to
\begin{equation}
\label{g"}
g''\,\cos\xi\, \simeq \ \frac{2m_U}{v}\ \,\frac{r}{\sin 2\beta}\,,
\end{equation}
with
\begin{equation}
\frac{1}{v}\,=\, 2^{1/4}\ G_F^{1/2}\,\simeq \ 4.06\times 10^{-6}\ \hbox{MeV}^{-1}\, .
\end{equation}
This leads from eq.~(\ref{gapm}) to
\begin{equation} \!
\label{gapm3}
  g_{A\pm}  \ \simeq  \ \displaystyle
   \frac{m_U}{2v}\ \,\frac{1 \pm \cos 2\beta}{\sin 2\beta}\ r
\ \simeq   \   \frac{m_U}{2v}
 \,\times\,\left\{
\begin{array}{cc}
 r\,\cot\beta\,,    &  \vspace{1mm}\\
  r\,\tan\beta\,,   & 
\end{array}\right.
\end{equation}
valid for supersymmetric or type~II models.
Similarly, we can derive from eq.~(\ref{gapm2}) the axial couplings for type~I models,
\begin{equation}
\label{gapm4}
  g_{A\pm}  \  \simeq \ \displaystyle 
\mp  \ \frac{m_U}{2v}\ r\ \tan\beta \ .
 \end{equation}
In both cases, we have
\begin{equation}
\label{ga4}
 g_{A\,d,e}\,=\,g_{A-}\,\simeq\,2.03 \times 10^{-6}\  m_U(\hbox{MeV}) \,\ r\ \tan\beta\ , 
\end{equation}
for down-type quarks and leptons.
This expression provides a general relation between $\varepsilon_A$, $m_U$, $\tan\beta$, and the invisibility parameter $r=\cos\theta_A$,
\begin{equation}
\label{epsilon}
\!\!\!\framebox[7.8cm]{\rule[-.3cm]{0cm}{8mm} \hbox{$ \displaystyle
  \varepsilon_{A\,d,e} \equiv \frac{g_{A\,d,e}}{e}\,\simeq\,6.7 \times 10^{-6}\  m_U(\hbox{MeV})\,\  r\,\tan\beta\,.$}}
\end{equation}

\vspace{2mm}

We then get from eqs.\,(\ref{gapm},\ref{gpga},\ref{mu},\ref{gapm3}) the equivalent pseudo\-scalar couplings
\begin{equation}
 g_{p\pm}     = \displaystyle  g_{A\pm}\,\frac{2m_{q,l}}{m_U}\, \simeq \ 4.06 \times 10^{-6}\, m_{q,l}(\hbox{MeV})\times \ \left\{
 \begin{array}{cc}
 r\,\cot\beta\,,    & \vspace{1mm} \\
r\,\tan\beta \,.  & 
\end{array}\right.
\end{equation}
This reconstructs precisely the couplings (\ref{gpax}) of the axion-like pseudoscalar (\ref{inva}) to quarks and leptons, 
the same ones as for a standard axion, multiplied by the invisibility factor $r=\cos\theta_A$ in eq.~({\ref{rinv}}).
In a similar way, we get from eq.~(\ref{gapm4}) the effective pseudoscalar couplings for type-I models,
\begin{equation}
 g_{p\pm}     = \displaystyle\,  g_{A\pm}\,\frac{2m_{q,l}}{m_U}\, = \ \mp\ \, 
 \frac{m_{q,l}}{v}\ r\,\tan\beta \,,
\end{equation}
which reconstruct precisely the Yukawa couplings of the axion-like pseudoscalar $a$ in eq.~(\ref{inva}), originating this time from the \,Im $h_1^0$ contribution to $a$, proportional to $r\,\sin\beta$.
\vspace{-2mm}

The fact that the light spin-1 $U$ boson in a longitudinal polarisation state gets produced or interacts much as a quasi-invisible axion-like pseudoscalar has been discussed and verified explicitly for the decays $\psi \to \gamma\,U\,,\ \Upsilon\to \gamma\,U\,,\ e^+e^-  \to \gamma\,U\,$, and for the $U$ boson contribution to the anomalous magnetic moment of the muon~\cite{Fayet:1980rr,Fayet:1981rp}. 
\vspace{2mm}

Moreover, when the couplings $g''$ and $g_A$ are expressed proportionately to $m_U$ and $r$ as in eqs.\,(\ref{g"},\ref{gapm3},\ref{gapm4}),
{\it all three polarisation states of a $U$ boson of a given mass $m_U$ decouple in the limit of small $r$}, for which the extra-$U(1)$ symmetry gets broken at a high scale.

\vspace{2mm}

\subsection{\boldmath Lifetime and decay length for an axial $U\!$ boson}

The partial lifetime for a $U$ boson decaying into a $f\bar f$ pair is
\begin{equation}
\label{gamma}
   \Gamma (U \to f \bar f)\,= \,\frac{1}{12\pi}\, \left[\,g_{Af}^2 \,\beta_f^3+\,g_{V\!f}^2\, (\,\hbox{$\frac{3}{2}$}\,\beta_f - \hbox{$\frac{1}{2}$}\,\beta_f^3)\,\right] \,m_U\,,
\end{equation}
where $\beta_f=v_f/c = (1-4m_f^2/m_U^2)^{1/2}$.
For purely axial couplings, the partial lifetime for leptonic decays, derived from eq.\,(\ref{epsilon}), can be expressed as
\begin{equation}
\label{tauee2}
\begin{array}{ccc}
   \tau_{ee}&\simeq&\displaystyle \frac{6\times 10^{-9}\,\hbox{s}}{r^2\,\tan^2\!\beta\ m_U\hbox{(MeV)}^3\ \beta_e^3}
   \vspace{2mm}\\
    &\simeq &\displaystyle\frac{2.7 \times 10^{-7}\ \hbox{s}}{m_U\hbox{(MeV)}\,(\varepsilon_{Ae}/10^{-6})^2\ \beta_e^3}\ ,
    \end{array}
\end{equation}
where we defined $g_A = \varepsilon_A e$, with similar expressions for $\tau_{\mu\mu}$ and $\tau_{\tau\tau}$. For invisible decays into neutrino pairs within supersymmetric or type II models, as indicated by eq.~(\ref{gapm3}), the decay lifetime can be approximated as
\begin{equation}
\label{taununu}
   \tau_{\nu\bar\nu}\,\simeq \,\frac{4\times 10^{-9}\,\hbox{s}}{r^2\,\cot^2\!\beta\ m_U\hbox{(MeV)}^3}
    \,\simeq \,\frac{1.8 \times 10^{-7}\ \hbox{s}}{m_U\hbox{(MeV)}\,(\varepsilon_{A\nu}/10^{-6})^2}\ .
\end{equation}
For type I 2HDMs, we replace $\cot^2\beta$ with $\tan^2\beta$ in the expression for the partial lifetime for invisible decays into neutrinos.
\vspace{2mm}

From now on, we shall focus on the most simple case of a purely axially coupled boson, where the $U$ directly gauges the axial symmetry $U(1)_A$, and there is no mixing with the $Z$ boson~\cite{Fayet:1980ad, Fayet:1980rr}. This situation naturally arises when $F= F_A$, meaning that $\gamma_A = 1$ and all other $\gamma$ values are zero in eq.\,(\ref{F}). 
We also set $F_1=F_2=1$ and $v_1=v_2$ or $\tan\beta=1$, which ensures no mixing with the $Z$ boson, according to eqs.~(\ref{m2},\ref{xi}).
The $U$ is axially coupled in a universal way to all standard model fermions (aside from neutrinos), with an axial coupling expressed from eq.\,(\ref{dcov}) as
$g_A = g''/4\,$ with 
\begin{equation}
\label{muu}
m_U^2\,=\,\frac{g''^2}{4\ }\,(v^2+F_\sigma^2\,w^2)\,=\,\frac{g''^2v^2}{4\,r^2},
\end{equation}
so that 
\begin{equation}
\label{epsaa}
\varepsilon_{A} \,=\, \frac{g_A}{e}  \,=\, \frac{g''}{4\,e}\,\simeq \,6.7 \times 10^{-6}\ m_U\hbox{(MeV)}\ r\,,
\end{equation}
as it was shown in eqs.\,(\ref{mu0},\ref{ga4},\ref{epsilon}).
\vspace{2mm}

A $U$ boson at least slightly heavier than $\simeq 1$ MeV but lighter than $2\,m_\mu$
would decay into $e^+e^-$ pairs about 40~\% of the time, and into neutrino pairs the remaining 60~\%. Its lifetime can be approximated by
\begin{equation}
\label{tau}
   \tau\,\simeq \,\frac{2.4 \times 10^{-9}\,\hbox{s}}{r^2\,m_U\hbox{(MeV)}^3}
    \,\simeq \,\frac{1.08 \times 10^{-7}\ \hbox{s}}{m_U\hbox{(MeV)}\,(\varepsilon_A/10^{-6})^2}\ .
\end{equation}
When the boson is ultrarelativistic, its decay length can be estimated as
 \begin{equation}
\label{l}
\begin{array}{ccl}
  l\,=\,\beta\gamma\,c\,\tau\,&\simeq & \displaystyle \,\frac{E\text{(MeV)}}{r^2\,m_U\text{(MeV)}^4}\ \times \,0.72\ \text{m} \vspace{2mm}\\
     &\simeq & \,\displaystyle
   \frac{E\text{(MeV)}}{m_U\text{(MeV)}^2\,(\varepsilon_A/10^{-6})^2}
   \times
   32.4 \ \text{m}\,.
\end{array}
\end{equation}

\begin{figure}[h!]
\includegraphics[width=0.5\textwidth]{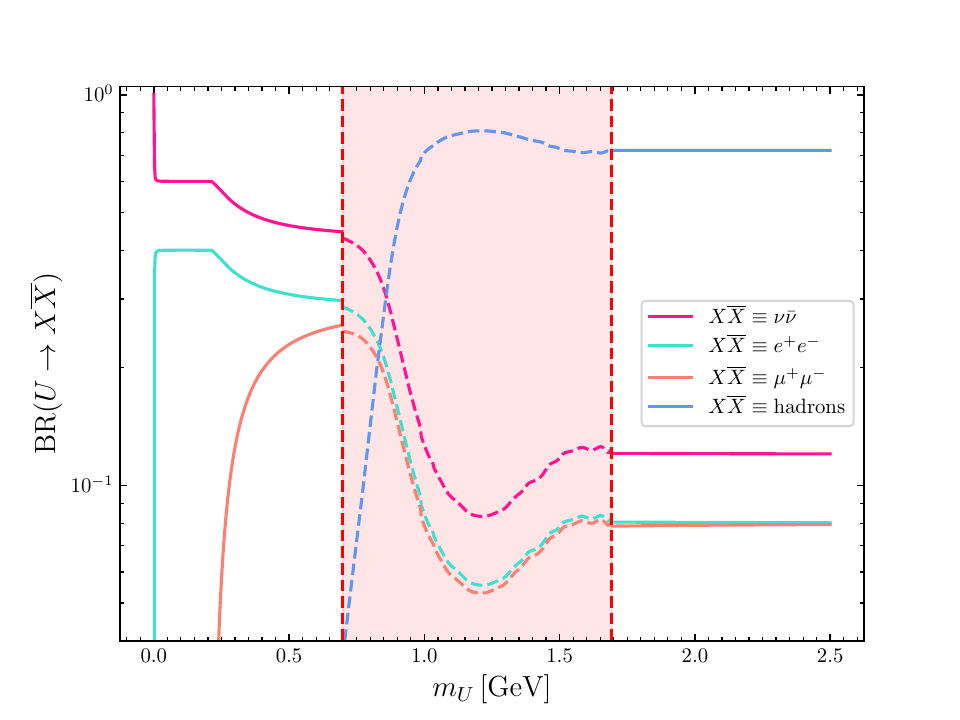}
\caption{Branching ratios of an axially-coupled $U$ boson, as a function of its mass. If $m_U$ is  slightly above $1 \MeV$ but less than $2m_{\mu}$, it decays 60\,\% of the time into neutrino pairs and 40\,\% into $e^+e^-$ pairs. A $U$ boson heavier than $1.5 \,\GeV$ tends to decay mostly into hadrons (72\,\%), with the remaining decays into neutrino pairs (12\,\%), electron-positron pairs (8\,\%), and muon pairs (8\,\%).
In the intermediate region between the dashed lines, we used Ref.\,\cite{Baruch:2022esd} to evaluate the partial decay width into hadrons.}
\label{fig:BR_U}
\end{figure}

   \vspace{2mm}
For $m_U \gtrsim 1.5 \,\GeV$, the $U$ boson would decay predominantly into $u\bar u,\ d\bar d$ and $s\bar s$ pairs, with partial decay widths 
\begin{equation}
\Gamma(U \rightarrow q \bar{q}) \,\simeq\, \alpha\, m_U \ \varepsilon_A^2 \,\beta_q^{3}\,.
\end{equation}
The partial decay width into hadrons for $m_U$ in the intermediate region, starting from $0.4$ to $0.7\, \GeV$ up to about $1.5 \, \GeV$, is determined as in ref.\,\cite{Baruch:2022esd}.
The resulting branching ratios are shown in fig.~\ref{fig:BR_U}.

   \vspace{2mm}

We shall now consider the possible production, decay and detection of new light gauge bosons in a muon beam dump experiment, verifying that longitudinally-polarised bosons get produced as axion-like pseudoscalars, and comparing the effects of axial and vector couplings in the production process of transversely-polarised ones.

\section{Beam dump experiments}
\label{sec:beam_dump}

\subsection{Production cross section}
Beam dump experiments are high-intensity facilities that aim to search for and measure the properties of elusive particles. They dump a collimated and mono-energetic beam of high-energy particles, typically protons or electrons~\cite{Coteus:1979wh,Jacques:1979vj, Bjorken:1988as, Bossi:2012uwa, Battaglieri:2017aum, Fabbrichesi:2020wbt, Graham:2021ggy, Antel:2023hkf}, into a dense block of heavy material placed shortly before a shielding (for an overview of the existing proton and electron beam dump experiments, see the comprehensive reviews in \cite{Fabbrichesi:2020wbt, Lagouri:2022ier, Graham:2021ggy}). This reduces the background of conventional leptons from the decays of known long-lived particles ($\pi$, $K$, $\Lambda$, ...) by hadron absorption in the dump and the shield. This feature enables the search for penetrating stable or quasi-stable particles, produced directly in the interactions with nuclei or from the decays of short-lived particles. 

\vspace{2mm}

Proton beam dump experiments \cite{Jacques:1979vj,Coteus:1979wh} have been used very early to get constraints on a new light gauge boson, with very weak axial couplings, that could be produced very much as an axion-like particle \cite{Fayet:1980rr}.
Without an additional dark singlet to contribute to $m_U$, the invisibility parameter has its maximum value at $r = \cos\theta_A = 1$, setting the axial coupling at $g_A = (g/4) \times m_U/m_W \simeq 2\times 10^{-6}\,m_U \, (\MeV)$, for $\tan \beta = 1$. Axial $U$  bosons with masses just over $1\, \MeV$ up to about $7 \,\MeV$ were then ruled out by early beam dump experiments, unlike heavier bosons which would stay undetected due to their too short decay lengths~\cite{Fayet:1980rr}.

\vspace{1.5mm}

\vspace{1.5mm}
The possibility of muon beam dump experiments has recently emerged~\cite{Cesarotti:2022ttv, Cesarotti:2023sje}, and it could complement the discovery potential of proton or electron beam dump experiments.
This is especially promising due to the stronger effective pseudoscalar coupling of the $U$ boson to muons compared to electrons, by a factor of $m_\mu/m_e$, as shown in eq.\,(\ref{gpga}).
We shall thus study the forward production of the new spin-1 gauge boson through muon bremsstrahlung $\mu + N \rightarrow \mu + U + X$, with the $U$ boson emitted from an incoming or outcoming muon. For a $U$ boson that is purely axially coupled, the indirect production via meson decay is significantly suppressed. Charge conjugation forbids the decays of $\pi_0$ and $\eta$ into $\gamma\, U$, and their decays into $UU$ are usually negligible due to the small coupling $g''$. 
\vspace{2mm}

Let us consider a muon with mass $m_\mu$, initial four-momentum $p$ and energy $E_0$. We denote the four-momentum of the emitted $U$ boson by $k$, with $x = E_U / E_0$ representing the fraction of the incoming energy carried away by the $U$. 
In the reference frame of the incoming muon, the rapidly moving target ``atom'' generates a cloud of virtual photons. The muon effectively interacts with it to emit a $U$ boson. Although these exchanged photons are spacelike, their virtuality is relatively small compared to other relevant invariants, resulting in the interaction between the muon and the target being predominantly driven by their transverse polarisations states~\cite{Bjorken:2009mm}. In the
one-photon-exchange process, the target particle,
viewed in the frame where it moves quickly in the opposite direction of the incident particle, behaves very much
as a beam of quasi-real photons produced by
the incoming charged lepton after it passes through a target of equivalent
thickness $\alpha \,\chi/\pi$ radiation length~\cite{Kim:1973he}.

\vspace{2mm}
This allows us to use the so-called Fermi-Weiszäcker-Williams (FWW) approximation~\cite{Fermi:1924tc, vonWeizsacker:1934nji, Williams:1934ad} and relate the full scattering process $\mu (p) + N(P_i) \rightarrow \mu (p') + N'(P_f)+ U(k)$ with the $2 \rightarrow2$ process $\mu (p) + \gamma(q) \rightarrow \mu (p') + U(k)$  evaluated at minimum virtuality $t_\text{min} = (-\,q^2)_\text{min}$, where we defined $q=P_f -P_i$, and the metric signature as $(+,-,-,-)$. The cross section in the laboratory frame may then be expressed as \cite{Bjorken:2009mm, Kim:1973he, Tsai:1973py, Tsai:1986tx}:
\vspace{-2mm}

\begin{equation}
\begin{array}{c}
\displaystyle \frac{d\sigma(p+P_i \rightarrow p' + k + P_f)}{d E_{U} d\cos \theta_{U}} \ \simeq \ \frac{\alpha \,\chi}{\pi}\ \frac{E_0\, x\, \beta_{U}}{1-x} \qquad \vspace{3mm}\\ [3mm]
\hspace{6mm}\displaystyle\times\ \,\frac{d\sigma(p + q \rightarrow p' + k)}{d(p\cdot k)}\ \bigg|_{\,t=t_{\rm min}},\ 
\end{array}
\label{eq:WWapprox}
\end{equation}
where
$\theta_{U}$ is the angle of emission, $\beta_{U}= \sqrt{1-m_{U}^2/E_{U}^2}$,
 and $\alpha=e^2/4\pi$ is the fine structure constant. This general expression holds independently of the type of couplings of the emitted $U$ boson. 
\vspace{0mm}
The FWW approximation is applicable when beam particles and emitted particles are highly relativistic and collinear, i.e.~when~\cite{Bjorken:2009mm}
\begin{equation}
\frac{m_\mu}{E_0}\,, \ \ \frac{m_{U}}{xE_0}\,, \ \ \theta_{U} \ll 1\ .
\end{equation}

The quasi-real photon flux is parameterised by the equivalent radiator thickness $\alpha \,\chi/\pi$, where~\cite{Kim:1973he, Tsai:1973py, Bjorken:2009mm}
\begin{equation}
\chi \,=\, \int_{t_\text{min}}^{t_\text{max}} dt\ \frac{t-t_\text{min}}{t^2}\ G_2(t)\,.
\label{eq:chi}
\end{equation}
Here $G_2(t) = G_{2,\rm el}(t) + G_{2,\rm in}(t)$ 
takes into account both atomic and nuclear contributions, as well as elastic and inelastic effects. Assuming that the cross section is dominantly collinear with $x$ close
to 1, the integration limits in eq.\,(\ref{eq:chi}) can be set to $t_{\rm min}=(m_U^2/2E_0)^2$ and $t_{\rm max}=$ $m_U^2+m_\mu^2$~\cite{Liu:2017htz}. For heavy targets such as lead, we have verified that the elastic contribution is typically more important, so that $G_2(t) \simeq G_{2,\rm el}(t)$, and it can be expressed as~\cite{Bjorken:2009mm}

\vspace{-5mm}

\begin{equation}
G_{2,\rm el}(t) = \left(\frac{a^2 t}{1 + a^2 t}\right)^2 \left(\frac{1}{1 + t/d}\right)^2 Z^2.
\label{eq:form_factor}
\end{equation}
\vspace{1mm}

The first factor, vanishing at $t=0$, corresponds to the effect of the elastic atomic form factor and characterises the screening of the nucleus potential by electrons at larger distances, in terms of the radius $a \simeq 111 \ Z^{-1/3}/m_e$. The second factor represents the effect of the finite size of the nucleus, and corresponds to the elastic nuclear form factor, in terms of $d \simeq 0.164  \ \text{GeV}^2 $ $A^{-2/3}$. 
The radiator thickness parameter $\chi$, proportional to $Z^2$, decreases when the $U$ boson mass increases as a result of the suppression in the form factor due to the finite size of the nucleus.

\vspace{2mm}

After integration over the angular dependence, the differential cross section in $x$ is~\cite{Liu:2017htz}
\begin{align}
&\!\!\!\!\frac{d\sigma}{dx}=\, 2\,\varepsilon_A^2\,\alpha^3 \, x \, \chi  \, \left[\ \frac{m_\mu^2\,x\,(2-x)^2-2\,(3-3x+x^2)\,\tilde{u}_{max}}{3\,x\,\tilde{u}_{max}^2} \right. \nonumber \vspace{1mm}\\
&\hspace{35mm}\left.
+\ \frac{2\,m_\mu^2\,(1-x)}{\tilde{u}_{max}\,(\tilde{u}_{max}+m_\mu^2x)}\ \right]\,,
\end{align}
where $\tilde{u}_{max}=-\,m_U^2\,(1-x)/x -m_\mu^2 x$. 
\vspace{2mm}

In the limit in which $m_U/m_{\mu}$ goes to zero, this expression is dominated by the last term in the bracket, so that 
\begin{equation}
\label{approximation}
\frac{d\sigma}{dx}\, \simeq\, \frac{4\,\varepsilon_A^2\,\alpha^3 \, x \, \chi}{m_U^2}\ .
\end{equation}
Not surprisingly, we get a production cross section proportional to $g_A^2/m_U^2$, as expected from the fact that in the small mass limit, an axially coupled 
$U$ boson is produced very much as an axion-like pseudoscalar, with an effective pseudoscalar coupling to the muon $g_P = g_A \times 2 m_\mu/m_U$ (see eq.\,(\ref{gpga})).
This greatly differs from the production of a light $U$ boson with vector couplings, which does not exhibit such characteristics when its mass is small.

\subsection{\boldmath \vspace{1mm} Expected number of events }

Let us evaluate the number of events which may be observed from the decays of $U$ bosons into $e^+e^-$ or $\mu^+\mu^-$ pairs. 
Let $N_\mu$ be the number of incoming muons with energy $E_0$ hitting a target characterised by length $L_T$, mass density $\rho$, and mass $m_T$ for each of its constituents, resulting in a surface density of $\rho\,L_T/m_T$. Then we can calculate the total number of $U$ bosons produced, each with an energy $x\,E_0$, directed longitudinally towards the detector, using the following expression
\begin{equation}
    N_U \,\simeq \,N_\mu\ \frac{\rho\,L_T}{m_T}\,\int\,\frac{d\sigma}{dx}\ dx\,.
\end{equation}
The integration limits have been set to $x_{\rm min}=m_U/E_0$ and $x_{\rm max} \simeq 1$.
We have assumed for simplicity negligible radiative energy losses of the incoming muons within the target.
\vspace{2mm}

Each one of the $U$ bosons produced has a decay length
\begin{equation}
    l_U\,\propto \,\frac{x\,E_0}{m_U^2\,\varepsilon_A^2}\,
\end{equation}
proportional to its energy  $x\,E_0$, as shown in eq.\,(\ref{l}).
To calculate the expected number of events, we must consider the average survival probability of a $U$ boson, produced at any point $z$ within the target length $L_T$, given by
\begin{equation}
\langle\, e^{-(L_T-z)/l_U}\, \rangle \,=\, \frac{l_U}{L_T}\  (1 - e^{-L_T/l_U})\,.
\end{equation}
This factor accounts for the $U$ boson likelihood of not decaying before exiting the target. We then multiply it by the probability  $e^{-L_{\text{sh}}/l_U}$ of the boson surviving after the shielding and by the probability $( 1 - e^{-L_{\text{dec}}/l_U} )$ of it decaying within the decay region. For a facility that detects $e^+e^-$ or $\mu^+\mu^-$ decays, the expected total number of detected events is determined by
\begin{align}
\label{nev}
    N_{\rm events} \,\simeq \,N_\mu\ \frac{\rho}{m_T}\,\left[\ \int\,\frac{d\sigma}{dx}\ \,\l_U\, (1-e^{-L_T/l_U}) \ e^{-L_{\rm sh}/l_U}\
    \right.
    \\ \nonumber
    \times\,\left. (1-e^{-L_{\rm dec}/l_U})\ dx\ \right]
    \,(B_{ee}+B_{\mu\mu})\,P_{\rm det}\,,
\end{align}
where $P_{\rm det}$ stands for the average detection probability.

\vspace{2mm}
When the interactions are extremely weak, resulting in decay lengths exceeding the experiment dimensions, the majority of the produced $U$ bosons exit undetected. The expression above further simplifies into
\begin{equation}
\label{nev2}
    N_{\rm events} \,\simeq \,N_\mu\ \frac{\rho\,L_T\,L_{\rm dec}}{m_T}\,\left[\,\int\frac{d\sigma}{dx}\,  
   \frac{1}{l_U}\, dx \,\right] \!
    (B_{ee}+B_{\mu\mu})\,P_{\rm det}\,.
\end{equation}
Considering that a very light ultrarelativistic $U$ boson is produced as an axion-like pseudoscalar proportionally to $\varepsilon_A^2/m_U^2$, it follows that
\begin{equation}
\label{nepsA}
N_{\rm events}\,\propto\,\frac{\varepsilon_A^2}{m_U^2\ l_U}\,\propto \ \varepsilon_A^4\,.
\end{equation}
For very weakly coupled axial bosons, an upper limit on the number of observed events will directly lead to a limit on $\varepsilon_A$,
typically 
\begin{equation}
 \varepsilon_A\ <\ 10^{-7},
\end{equation}
for a number of incoming muons $N_\mu \approx 10^{20}$. This limit is essentially independent of $m_U$ as long as it is smaller than $2\,m_\mu$,  as illustrated in fig.~\ref{fig:ExclusionRegion}.

\vspace{2mm}

For moderately coupled $U$ bosons, which can be sufficiently produced to be observable, the situation changes. The key factor is then the exponential term $e^{-L_{\rm sh}/l_U}$, which is typically around $10^{-5}$ for a 10 m shielding and a decay length $l_U$ of slightly below 1 m. This exponential term decreases significantly for smaller $l_U$ values. Consequently, $U$ bosons can be detected within the $\varepsilon_A$ range of approximately $10^{-4}$ to $10^{-6}$, also depending on the considered mass $m_U$. When $l_U$ is somewhat smaller than both $L_T$ and $L_{\rm dec}$, the number of events simplifies to
\begin{align}
\label{nev3}
    N_{\rm events} \,\simeq \,N_\mu\ \frac{\rho}{m_T}\,\left[\ \int\,\frac{d\sigma}{dx}\ \,\l_U\,  \ e^{-L_{\rm sh}/l_U}\
   \ dx\ \right]
    \\ \nonumber
    \,(B_{ee}+B_{\mu\mu})\,P_{\rm det}\,.
\end{align}
The decay length $l_U$ behaves as $(m_U\,\varepsilon_A)^{-2}$, as shown in eq.~(\ref{l}). Consequently, the limit on $\varepsilon_A$ is essentially inversely proportional to $m_U$. For $m_U$ values less than 1 GeV, in the absence of detected events, the expected constraint on $\varepsilon_A$ can be approximated as
\begin{equation}
\label{epsamu}
\varepsilon_A >\  10^{-4} \times \frac{100 \ \rm MeV}{m_U}\,,
\end{equation}
as we shall see in fig.~\ref{fig:ExclusionRegion}.

\subsection{Expected exclusion region}

\label{s:Results}
While muon beam dump experiments can serve as powerful means to investigate smaller values of the couplings of new gauge bosons~\cite{Cesarotti:2022ttv,Cesarotti:2023sje}, it is noteworthy that numerous characteristics defining the form of the exclusion region are not unique to muon beam dumps but are shared with proton and electron beam dumps. In the present section, we will discuss the common and distinctive features of the proposed muon beam dump facility in the context of probing new light weakly coupled axial forces. While previous studies~\cite{Cesarotti:2023sje} have examined light axially coupled forces in muon beam dump experiments with a focus on muonphilic couplings, here we focus on a universal coupling of the boson to all standard model quarks and leptons.
\vspace{2mm}

Muon beam dump experiments present a unique opportunity for probing such new interactions, particularly when compared to other beam dump facilities. This stems from the fact that a light $U$ boson with axial couplings undergoes interactions that are significantly enhanced as shown  in eq.\,(\ref{gpga}). For muons, this enhancement factor is about 200 times larger than what can be achieved for electrons in an electron beam dump experiment. 

\vspace{2mm}

For the experimental set-up, we will assume an expected number of muons on target of  $N_{\mu} = 10^{20}$. This estimation aligns with prior studies~\cite{Cesarotti:2022ttv}, utilising the MAP design parameters~\cite{Delahaye:2013jla, Neuffer:2018yof}. Regarding the energy of the incoming muon beam, we will adopt $E_0 = 1.5 \,\TeV$, corresponding to a $3\, \TeV$ collider, which is a standard benchmark in the literature on muon collider proposals~\cite{Delahaye:2019omf}. As an illustrative example,  we consider a target length of $L_T = 10 \,\meters$, a decay region length of $L_\text{dec} = 100\, \meters$, and a shielding extent of $L_\text{sh} = 10\, \meters$. We assume lead to be the target material.

\begin{figure}[h!]
\includegraphics[width=0.47\textwidth]{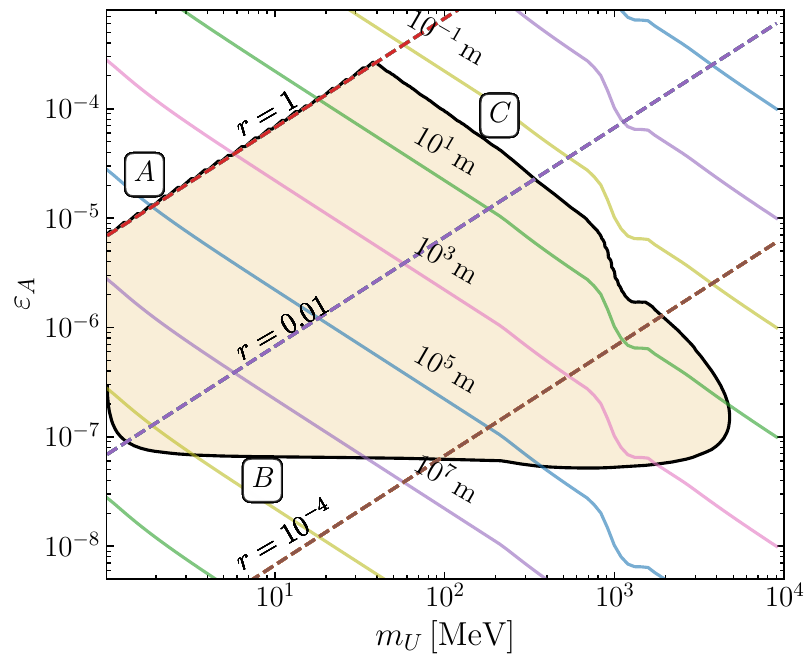}
\caption{Contour line corresponding to three signal events detected, for $N_\mu = 10^{20}$ and $E_0 = 1.5 \,\TeV$ (black), limiting the exclusion area (beige) in case no events are observed. Coloured solid lines indicate constant decay lengths $l_U$, evaluated using (\ref{l}) for $x=1$. Coloured dashed lines correspond to a constant invisibility parameter $r\leq 1$. The different boundaries of the exclusion region are labelled by $\mathbf{A}$, $\mathbf{B}$ and $\mathbf{C}$,
and are discussed in the text.}
\label{fig:ExclusionRegion}
\end{figure}

\vspace{2mm}
In fig.~\ref{fig:ExclusionRegion}, the black line encloses the exclusion region under the assumption of no signal events being observed, in which we used $N_{\rm events}=3$ to set the black contour. The solid lines in color represent lines of constant decay length (evaluated for $U$ bosons of energy close to $E_0$), whereas the dashed lines represent lines of constant invisibility parameter $r$.

\vspace{2mm}

The top-left boundary of the exclusion region, labelled by $\bf{A}$, arises when only the two scalar doublets contribute to the gauge boson mass and no singlet is present, corresponding to $r=1$ with $\tan\beta =1$ and $m_U=m_U^{0}= g''v/2$ in eqs.\,(\ref{muu},\ref{epsaa}).  Given a fixed coupling constant $\varepsilon_A\,e$ and fixed value of $\tan \beta$, the mass of the $U$ boson cannot be arbitrarily small, since two BEH doublets are required to gauge the axial symmetry, imposing a minimum mass limit. An increase in the dark singlet v.e.v. results in reduced values of the invisibility parameter $r$, leading to a heavier $U$ boson at constant coupling values, i.e.~constant $\varepsilon_A$, as we see for $r=0.01$ and $r=10^{-4}$.
\vspace{2mm}

The bottom boundary of the black contour, indicated by $\mathbf{B}$, is largely unaffected by variations in mass. For feeble enough interactions, the decay length $l_U$ becomes large with the number of expected events, which in general is given by eq.\,(\ref{nev}), reducing to the simpler expression (\ref{nev2}). Considering $d\sigma/dx$ for small values of $m_U$ (see eq.\,(\ref{approximation})), $l_U\simeq xE_0/m_U \times \tau_U$ and $\tau_U^{-1}= \Gamma_U \simeq \frac13 \,\varepsilon_A^2\,\alpha\,m_U /B_{ee}$, one gets
\begin{equation}
\label{nev4}
N_{\rm events} \,\simeq\, \frac{4\,\alpha^4\,N_\mu\,\rho\,\chi \,L_T\, L_{\rm dec} }{3 \,E_0 \,m_T} \ \,\varepsilon_A^4\,P_{\mathrm det}\,,
\end{equation}
which has lost its leading dependence on $m_U$.  For values of $\varepsilon_A$ lower than those specified by the boundary $\mathbf{B}$, highly long-lived $U$ bosons pass through the detector without interacting with it. 

\vspace{2mm}

The top-right boundary, labelled by $\textbf{C}$, depends on the detector geometry and the particle decay length. This boundary is nearly parallel to the lines of constant decay length. The number of expected signal events for $U$ bosons with a short decay length is suppressed by the middle exponential term in eq.\,(\ref{nev}), which then reduces to the simpler form in eq.\,(\ref{nev3}). Therefore, we cannot constrain $U$ bosons with $l_U$ smaller than a few decimeters with these choices for the experimental set-up ($L_T = 10\meters$ and $L_\text{sh} = 10\meters$), since they decay well before reaching the detector.
The allowed region for $\varepsilon_A$ essentially indicates that the decay length $l_U$ (evaluated for the highest-energy $U$ bosons where $x$ is nearly 1) must be shorter than an upper boundary, which is just under 1 m. Since $l_U$ behaves as $m_U^{-2}\,\varepsilon_A^{-2}$, this results in a constraint on $\varepsilon_A$ that is inversely proportional to $m_U$, as shown in eq.~(\ref{epsamu}).
\vspace{2mm}

The shape of the black contour remains largely independent of the specific details of the beam dump experiment considered, for the reasons outlined above. The size of the exclusion region, however, is sensitive to the specific parameters of each experiment. For the previously mentioned set-up in a muon beam dump experiment, we could probe coupling strength values as low as $10^{-7}$ across a mass range from $1.1\, \MeV$ to $4.7 \,\GeV$.

\subsection{Comparison \vspace{1mm}between the axial, vector and pseudoscalar cases}

To compare the cases of pseudoscalar, axially and vectorially coupled particles, let us distinguish the longitudinal and transverse polarisation states of an axially coupled boson. Its production cross section can be decomposed as
\begin{equation}
    \sigma_{\text{prod}} (U) \,= \, \sigma_{\text{prod}} (U_{L}) + \sigma_{\text{prod}} (U_{T})\,.
\end{equation}
In the small $m_U$ limit (i.e.~for $m_U \lesssim$ 20 MeV), the first term, $\sigma_{\text{prod}} (U_{L})$, is dominant and it is similar to that of an axionlike pseudoscalar.
The second term, $\sigma_{\text{prod}} (U_{T})$, closely resembles that of a vectorially coupled particle when the mass of the muon $m_\mu$ can be neglected. 
In this case, the axial and vector couplings to the muon are virtually indistinguishable, with the production of the longitudinal polarization state being negligible. This equivalence becomes apparent when $m_U$ exceeds roughly twice the muon mass, $2m_\mu$.

\vspace{2mm}

We thus consider a pseudoscalar particle $U$ with couplings to standard model fermions
\begin{equation}
\label{coup}
 \varepsilon_P e \ U \, \bar f\gamma_5 f\,,
\end{equation}
and a vectorially coupled boson with couplings 
\begin{equation}
\label{coup2}
 \varepsilon_V e \ U_\mu\,\bar f \gamma^\mu f \,,
\end{equation}
choosing $\varepsilon_P=\varepsilon_A \times 2 m_{f}/{m_U}$ and $\varepsilon_V = \varepsilon_A$ for comparing the production cross sections.
\vspace{2mm}

\begin{figure}[h!]
\includegraphics[width=0.47\textwidth]{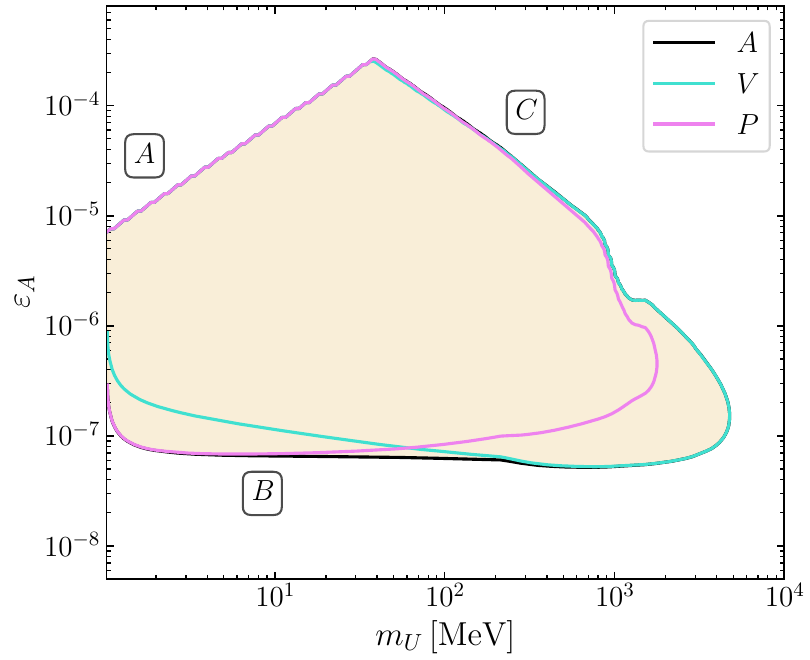}
\caption{Contour line corresponding to three signal events detected for $N_\mu = 10^{20}$ and $E_0 = 1.5 \,\TeV$ (black) and exclusion area assuming no events are observed (beige), for an axially coupled $U$ boson. The pink and green lines indicate the three signal event contours for an auxiliary pseudoscalar particle and a vectorially coupled $U$ boson, respectively. An axially-coupled boson, if sufficiently light, is produced as an axion-like pseudoscalar, while a heavier one is produced as a vectorially-coupled boson.}
\label{fig:Comparison}
\end{figure}

In the low mass and coupling region the interactions of an axially coupled gauge boson have enhanced interaction amplitudes, mimicking those of a pseudoscalar particle coupled as in (\ref{coup}). One should then expect an increased number of signal events as compared to a vectorially coupled gauge boson as in (\ref{coup2}). Fig.~\ref{fig:Comparison} compares the potentially excluded parameter regions for an auxiliary pseudoscalar (pink) and an auxiliary gauge boson with pure vector couplings (green), using the same experimental configuration as previously described for an axially-coupled boson. To facilitate the comparison, we have considered that the decay lengths of the pseudoscalar and of the vectorially coupled particles be the same as for an axially coupled one. This provides a simplified illustration for comparative purposes of the three production cross sections and does not represent a physically realistic situation for the pseudoscalar and vectorially coupled particles.

\vspace{2mm}

The differential production cross section for the pseudoscalar particle is expressed as~\cite{Liu:2017htz}
\begin{equation}
\frac{d\sigma}{dx}\, =\, \varepsilon_P^2 \,\alpha^3 \,x\,\chi \ \frac{m_\mu^2 x^2 - 2x \tilde{u}_{\text{max}}}{3\tilde{u}_{\text{max}}^2}\ .
\end{equation}
In the limit where $m_U\to 0$, we have $\tilde{u}_{\text{max}}\simeq -\,m_\mu^2\,x$, and
\begin{equation}
\frac{d\sigma}{dx}\, \simeq\, \frac{\varepsilon_P^2\,\alpha^3 \, x \, \chi}{m_\mu^2}\ .
\label{approximationps}
\end{equation}
We recover the same expression of the differential cross section as for an axially coupled boson in eq.\,(\ref{approximation}), 
with the correspondence  
$\varepsilon_P=\varepsilon_A \times 2 m_{f}/{m_U}$. The expected number of signal events $N_{\rm events}$ at low values of the auxiliary pseudoscalar particle mass and large decay length is still given by the same eq.\,(\ref{nev4}), so that the corresponding pink contour at the boundary {\bf B} of fig.\,\ref{fig:Comparison} coincides with the black one for a light axially coupled $U$ boson.
\vspace{2mm}

On the other hand, the differential production cross section of a vectorially coupled spin-1 gauge boson is expressed as in ref.~\cite{Liu:2017htz}.
When $m_U$ is very small, it is predominantly given by
\begin{equation}
\frac{d\sigma}{dx}\, \approx\, 2\,\varepsilon_V^2\,\alpha^3\, \chi\ \frac{3x^2 - 4x + 4}{3 m_\mu^2 x},
\end{equation}
which does not increase in the limit of vanishing mass at fixed coupling constant. For large decay lengths and small mass values, the number of signal events, predominantly involving soft $U$ bosons with small $x$ values, is estimated to be of the order of the leading term in $m_U$
\begin{equation}
N_{\rm events}^V \,\approx\, \frac{8 \,\alpha^4\, N_{\mu}\, \rho\, \chi \,L_T\,L_{\text{dec}}\,m_U}{9 \,m_{\mu}^2 \,m_T}\ \varepsilon_V^4\ P_{\rm det}\,,
\end{equation}
which decreases as the mass $m_U$ decreases. 
\vspace{-.5mm}
The upper limit on $\varepsilon_V$ in this region thus behaves roughly as $m_U^{-1/4}$, as depicted by the green line close to region {\bf B} of
fig.~\ref{fig:Comparison}.
This illustrates how the production of an axially coupled gauge boson is enhanced compared to a vectorially coupled one (green) at low mass and coupling values.

\vspace{2mm}

The excluded region for the  auxiliary pseudoscalar coincides with that for the axial vector (black) in the low mass regime, i.e.~for $m_U \lesssim \ 20 \,\MeV$, as a very light spin-1 axially coupled $U$ boson behaves as a pseudoscalar, with the same differential production cross section at leading order, given by eq.\,(\ref{approximation}).
On the other hand, in the higher mass regime ($m_U \gtrsim 100 \,\meV$), the exclusion region for the axially coupled case is essentially the same as for a gauge boson with pure vector couplings. This is due to the fact that, at higher masses, the dominant contributions to the production cross-section of an axially-coupled boson arise mainly from its transverse polarisations, which behave in the same way as those of a vectorially coupled particle.
\vspace{2mm}

As previously noted, the upper-right boundary $\mathbf{C}$ is mostly determined by the detector geometry and the particle decay length. Given our choice to use the same decay lengths, for the sake of comparing the production cross sections, all three situations lead to the same upper boundary in this region {\bf C}.
\vspace{2mm}

In a muon beam dump experiment, the correspondence between an axion-like particle and a $U$ boson becomes manifest when $m_U \ll 2m_\mu$. This results in a substantial difference in the expected numbers of signal events for an axially coupled boson, as compared to the case of pure vector couplings. Such a distinction becomes challenging within an electron beam dump setting, as the enhancement factor $2m_e/m_U$ is much smaller. A muon beam dump can then better serve to differentiate between situations involving the presence or absence of axial couplings.
\vspace{2mm}

\begin{figure}[h!]
\includegraphics[width=0.47\textwidth]{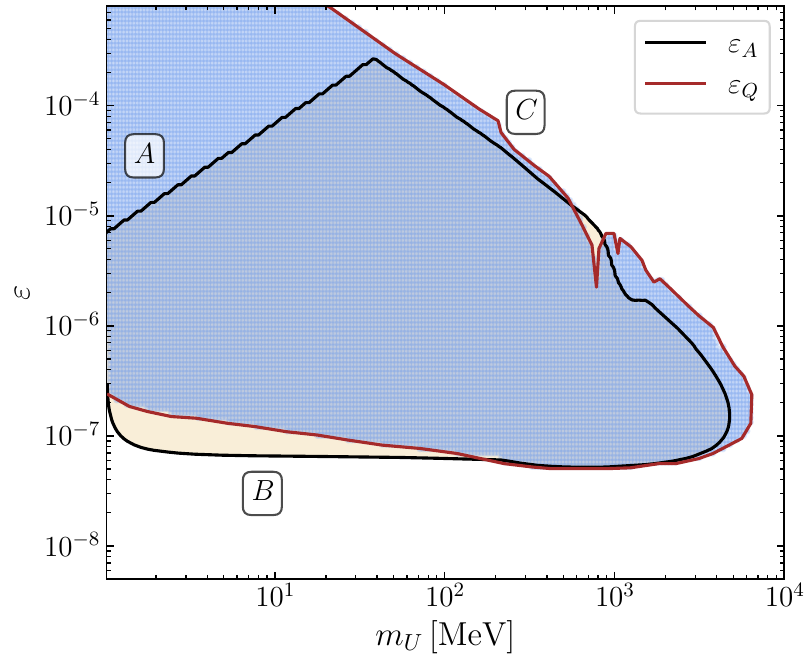}
\caption{Contour line corresponding to three signal events detected for $N_\mu = 10^{20}$ and $E_0 = 1.5 \,\TeV$ (black) and potential exclusion region assuming no events observed (beige), for an axially coupled $U$ boson. For the dark photon case, the red contour corresponds to three signal events for the same set-up, and the blue area denotes the exclusion region. The boundaries in region {\bf C} are principally sensitive to the decay lengths of the new bosons. The higher boundary for a pure dark photon, for $m_U< 2m_\mu$, comes from the fact that, with no decays into neutrinos it must have a coupling parameter $\varepsilon_V$ somewhat larger than the corresponding $\varepsilon_A$ for an axial boson, for a comparable decay length.
\vspace{3mm} }
\label{fig:Comparison2}
\end{figure}

The properties outlined above are further illustrated in fig.\,\ref{fig:Comparison2}, where we compare the case of a pure dark photon, which couples proportionally to the electric charge as expressed in eq.~(\ref{eq:DP_case}), with that of an axially coupled boson with universal fermionic couplings, as in eq.~(\ref{epsaa}).
Taking into account the actual lifetime of the dark photon has only a modest effects on the resulting contours, as seen by comparing figs.~\ref{fig:Comparison} and \ref{fig:Comparison2}.
The lifetime effect in the dark photon case is mainly apparent in the resonance region in fig.~\ref{fig:Comparison2}. This  effect is also visible in region {\bf C}, for which the boundaries are mostly sensitive to the decay lengths of the new bosons. The higher boundary for a pure dark photon for $m_U< 2m_\mu$ comes from the fact that, with no decays into neutrinos, it must have a coupling parameter $\varepsilon_V$ somewhat larger than the corresponding $\varepsilon_A$ for an axial boson, for a comparable decay length. Within boundary {\bf B}, the enhanced interactions with muons for low-mass axial bosons result in a larger exclusion region, compared to dark photons. In region {\bf A}, the $U$ boson mass cannot be arbitrarily small for a given value of $\varepsilon_A$, a constraint not applicable to dark photons or vectorially coupled bosons.

\vspace{2mm}

\section{Conclusions}
\label{sec:conclu}

We have presented a general formalism for new $U(1)$ interactions involving the weak hypercharge generator $Y$, the baryon and lepton numbers $B$ and $L_i$, and possibly, in the presence of a second Brout-Englert-Higgs doublet, an axial symmetry generator $F_A$. 
After mixing with the $Z$ boson, the resulting $U$ boson interpolates between a generalised dark photon, a dark $Z$ boson and an axially coupled gauge boson.
We have paid particular attention to its axial couplings, originating from the axial $F_A$ or associated with the mixing with the $Z$ boson.  We have shown how the scalar sector of the theory plays an essential role in the determination of these axial couplings, for instance, through the ratio $\tan\beta$ of the two doublet v.e.v.s~and the large v.e.v.~of an extra dark singlet. 
\vspace{2mm}

Our research highlights unique characteristics of these new axial interactions. For example, they lead to enhanced interactions of the longitudinally polarised state of the $U$ boson in the ultrarelativistic limit, making it behave very much as an axion-like particle, with effective pseudoscalar couplings to quarks and leptons $g_A \times 2m_f/m_U$.
This enhancement is particularly relevant for potential future beam dump experiments using muon beams, as the muon mass considerably enhances the effective coupling, proportional to $2m_\mu/m_U$, compared to analogous experiments with electrons.
\vspace{2mm}

Another distinctive feature of axial interactions is the way in which the parameters of the scalar sector influence the $U$ boson phenomenology. Notably, the mass of the $U$ boson, given a fixed axial coupling constant $\varepsilon_A \, e$, is subject to a lower bound due to the requirement of two Brout-Englert-Higgs doublets. This constraint, often overlooked, emerges inherently from a realistic model of axially coupled interactions.

\vspace{2mm}

We discussed in detail the shape of the expected beam dump exclusion or discovery region, and how it arises from the interplay between the $U$ boson interactions and the geometry of the experimental setup. For not-too-weakly interacting bosons (i.e.~with couplings $\approx 10^{-4}\ e$), the distance between the target and the detector plays a crucial role.
Bosons with relatively strong couplings are likely to decay too soon, never reaching the detector.  Conversely, those with extremely weak couplings tend to be nearly stable. In such cases, the limit on the axial coupling parameter $\varepsilon_A = g_A/e$ is typically on the order of $10^{-7}$ (for customary choices of the experimental set-up), and largely independent of $m_U$.
For lower values of $m_U$ the production cross section is dominated by the longitudinal polarisation state of the $U$ boson, produced much as an axion-like particle (see black and pink curves in fig.~\ref{fig:Comparison}). At higher masses ($m_{U} \gtrsim m_{\mu}$) on the other hand, the production 
corresponds almost exactly to that of a vectorially coupled boson (see figs.~\ref{fig:Comparison} and \ref{fig:Comparison2}). This illustrates how the interactions of a gauge boson with axial couplings are dominated by its longitudinal state at lower masses and by the transverse ones at higher masses, leading to a rather universal shape of the corresponding exclusion or discovery regions. Understanding how the parameter space that may be tested depends on the geometry of the apparatus and interactions of the new boson is also essential for the design and optimisation of future beam dump experiments.

\vspace{3mm}

\section*{Acknowledgments}
M.O.O.R. thanks Inar Timiryasov, Nikita Blinov and Pedro Cal for useful discussions. The work of M.O.O.R~is supported by the European Union's
Horizon 2020 research and innovation programme under grant agreement No 101002846, ERC CoG ``CosmoChart''.

\bibliography{refs}
\bibliographystyle{JHEP}

\end{document}